\documentclass[12pt]{article}
\usepackage[left=2cm,top=2cm,right=2cm,bottom=2cm,nohead,nofoot]{geometry}
\usepackage{graphicx}
\usepackage{setspace}
\usepackage[square]{natbib}
\usepackage{amssymb}
\usepackage{verbatim}
\usepackage[usenames,dvipsnames]{color}

\newcommand{\imloc}{}
\newcommand{\baseloc}{../../}

\begin{document}

\title{Valued Ties Tell Fewer Lies, II: Why Not To Dichotomize Network Edges With Bounded Outdegrees}
\author{Andrew C. Thomas\thanks{Visiting Assistant Professor, Department of Statistics, Carnegie Mellon University. Corresponding author: email \href{mailto:act@acthomas.ca}{act@acthomas.ca}. This work was supported by grant PO1-AG031093 from the NIA through the Christakis lab at Harvard Medical School and DARPA grant 21845-1-1130102 through the CMU Statistics Department.} \and Joseph K. Blitzstein\thanks{Assistant Professor, Department of Statistics, Harvard University.}}
\date{\today}

\maketitle

\begin{abstract}
Various methods have been proposed for creating a binary version of a complex network with valued ties. Rather than the default method of choosing a single threshold value about which to dichotomize, we consider a method of choosing the highest $k$ outbound arcs from each person and assigning a binary tie, as this has the advantage of minimizing the isolation of nodes that may otherwise be weakly connected. However, simulations and real data sets establish that this method is worse than the default thresholding method and should not be generally considered to deal with valued networks.
\end{abstract} 

\onehalfspacing

\section{Introduction}

When considering complex networked systems, there is a strong history of reducing the relative strengths of connections into a simpler, binary format. Much of this has to do with the history of the discipline and its roots in graph theory, though as the field of complex network analysis has grown into the analysis of many types of data, much of the software that exists for this analysis is still applicable only to binary connections. This has spawned a number of attempts to create binary versions of valued networks so as to perform some kind of analysis. A standard strategy for accomplishing this is to apply a uniform threshold to all ties in the set, set all higher values to ``one'' and all below to ``zero''. \citet{thomas2010vttfl} shows that even ``principled'' choices of threshold can differ wildly from each other, so that substantial amounts of information can be lost, and moreso than in other settings where dichotomization is practiced. (\citet{thomas2010vttfl} also contains a more thorough discussion of dichotomization of valued networks.)

One of the issues with a straight threshold is that a network that was originally fully connected may now have disconnected components, especially weakly connected nodes that now become singletons; this can lead to other cases of bias, including the notion that ``strong'' and ``weak'' ties have different functions in social networks \citep{granovetter1973swt}. If a binary version of a valued graph is accurate, it is likely that both of these kinds of ties must be respected in some way.

There is also the known problem that many network data sets are incomplete due to the friend-naming mechanism \citep{thomas2010cocisnpeaa}, which typically asks respondents to name up to a fixed number of contacts (who will themselves hopefully be in the network) and censors the rest from view. Assuming that all respondents' choices of friends are true, the consequences of this can be mitigated with a high limit on the number of potential names. 

Combining these factors leads to an intriguing possibility for dichotomizing the network: the artificial implementation of the ``name-$k$-friends'' strategy at the analysis stage. By deliberately choosing each individual's $k$ strongest ties as the binary network, and resymmetrizing if necessary, a binary version of the valued graph can be produced that may preserve whatever features of interest in the system are of interest to the investigator.

The procedure for a particular valued graph, whether from a real data set or simulated according to a model family, is as follows:

\begin{itemize}

\item Choose a feature, or set of features, of the valued graph to be preserved in the transformation. These can correspond to nodal characteristics like closeness or betweenness centrality, or global properties like diameter.

\item Select a ladder of maximal out-degrees to which nodes can be censored. For each outdegree value $k$, create a directed graph by selecting the top $k$ outbound arcs in value, censoring the rest and assigning the non-zero ties a value of 1. 

\item If the original graph is undirected, symmetrize the dichotomized directed graph to produce an undirected counterpart, by assigning an edge if either or both arcs is equal to 1.

\item For each maximal out-degree, calculate the relevant statistics for the dichotomized graph. Compare these to the valued graph and choose an optimal maximal out-degree for each statistic in the selection according to this statistic.

\end{itemize}


To investigate the effectiveness of this procedure, we first outline the simulation of networks from a broad generative model framework, as specified in \citet{thomas2010mshmfrd}. we then outline a number of features that can be preserved in the transformation of the valued graphs into binary counterparts, and demonstrate the method on a series of simulated and real examples. We show that this method performs worse on two real data sets than the standard thresholding procedure, and still maintains many of its flaws on simulated data, namely that networks with high heterogeneity on node degree are not well-preserved under the transformation.

\section{Simulation Models}

We use the Generalized Linear Model framework of \citet{thomas2010mshmfrd} to generate valued, undirected networks with various characteristics and properties. In particular, specify $Y_{ij}$ as the valued connection between nodes $i$ and $j$; in the simulation process, we consider only undirected graphs so that $Y_{ij} = Y_{ji}$.

The generation proceeds according to the following recipe (identical to that in \citet{thomas2010vttfl}): 

\begin{itemize}

\item Select a generative family from the GLM toolkit where edges have nonnegative value, such as 

\[ Y_{ij}|\mu_{ij} \sim \frac{1}{\mu_{ij}}Gamma(\mu^2_{ij}) \]

\noindent or

\[ Y_{ij}|\mu_{ij} \sim Poisson(\mu_{ij}). \]

The former will give a continuum of tie values, while the latter can produce graphs that contain explicit zeros.

\item Select a series of latent parameters that define $\mu_{ij}$:

\begin{itemize}

\item Sender/receiver effects $\alpha_i \sim N(0,\sigma_{\alpha}^2)$, where a larger $\sigma_{\alpha}$ yields more heterogeneity between nodes.

\item Latent geometric structure: nodes have positions $\vec{d}_i$, and a coefficient of distance vs. connectivity $\gamma$. For this investigation, nodes can lie equally spaced on a ring of unit radius, or in a single cloud from a bivariate normal distribution. 

\item Latent clusters. Each node is assigned membership in one of three clusters ($a_i=k$), and prefer links either within their cluster or with those nodes in other clusters with propensity $\lambda$. In this simulation, only one of clusters and geometry can be implemented at one time.

\item An assortative mixing factor $\chi$ equal to 0.5, 0 or -0.5 (the disassortative case), which determines whether degree itself is a factor in tie formation; in the assortative case, the popular individuals are more likely to form ties with with each other than otherwise expected, and likewise for the unpopular individuals.

\item The number of nodes in the system, in this case between 50 and 600 (limited upwards to make computation tractable).

\end{itemize}

All together, this gives an outcome parameter for a tie equal to

\begin{equation} \mu_{ij} = \alpha_i + \alpha_j + \chi \alpha_i \alpha_j - \gamma|\vec{d}_i - \vec{d}_j| + \lambda \mathbb{I}(a_i=a_j); \label{master-sim} \end{equation}

\noindent note that this is symmetric in $i$ and $j$. A list of all options for the above parameters is shown in Table \ref{threshold-sim-types}. 

To keep the parameter values positive, their values are bounded above zero with the transformation function $\mu_{pos} = f(\mu) = exp(\mu-1)\mathbb{I}(\mu<1) + \mu\mathbb{I}(\mu \geq 1)$, rather than setting all negative-parameter draws to zero; in execution, the difference is negligible compared to the magnitude of the other ties in the system.

\end{itemize}

\begin{table}
\begin{center}
\begin{tabular}{c|ccccccc}
Quantity & Values & & & & & \\
\hline
Nodes & 50 & 100 & 200 & \textit{300} & \textit{400} & \textit{500} & \textit{600} \\
Pop/Greg Signal & 0.1 & 0.5 & 1 & 2.5 & 10 & 100 \\
Geometry & None & Ring & Cloud & Cluster + & Cluster - & \\
Geo. Strength & 0.25 & 3 & & & & \\
Assortative Mixing & 0 & 0.5 & -0.5 & & & \\
Family & Gamma & Poisson & & & &
\end{tabular}
\end{center}
\caption{\label{threshold-sim-types} Simulation parameters to investigate the effects of censoring and dichotomizing valued networks by out-degree as an alternative to threshold-based dichotomization. Networks with 50, 100 and 200 nodes are simulated in the tests of geometric properties; larger networks (up to 600) are also generated for the linear model implementations.}
\end{table}

\section{Effect on Network Characteristics}

The statistical measures we consider are based on two families of distance measures on graphs:

\begin{itemize}

\item Geodesic measures, which are based on the shortest path distance $d(i,j)$ between two nodes $i$ and $j$ (see \citet{freeman1979csncc} for an excellent overview of these methods). The reciprocal of this is the closeness $1/d(i,j)$ which has the property that two nodes in separate components have zero closeness, rather than infinite distance.

\item Ohmic measures, which are based on the interpretation of social ties as resistors (or, more appropriately, conductors) in an electrical grid, so that the distance $d_{\Omega}(i,j)$ between two nodes $i$ and $j$ is equivalent to the resistance of the circuit formed by connecting nodes $i$ and $j$ (with symbol $1/G_{ij}^{eq}$, so that $G_{ij}^{eq}$ is the social equivalent of electrical conductance). The notion is useful in physical chemistry \citep{klein1993rd, brandes2005cmbcf} but is also finding new uses in complex network analysis due to its connections with random walks and eigenvalue decompositions \citep{newman2005mbcbrw}. \citet{thomas2009ocindac} gives a more thorough analysis of these measures and their comparisons with their geodesic equivalents; what is most relevant is that these are more sensitive to the total length of all paths that connect two points, to which geodesic measures, concerned only with the shortest single path, are largely indifferent.

\end{itemize}

For any network, valued or binary, there is a collection of graph statistics based on geodesic and Ohmic measures that apply to the individuals within. The choice of threshold affects the node statistics both in absolute terms and relative to each other, and the inherent uncertainty in the measurement of tie values suggests that these statistics vary between different iterations at the same threshold level, hence the increased reliability of using a number of replicates.

For this analysis, three measures of network centrality are considered:

\begin{itemize}

\item Harmonic geodesic closeness, $C_{1/C}(i) = \sum_j \left( \frac{1}{d(i,j)} + \frac{1}{d(j,i)} \right) $;

\item Ohmic closeness, $C_{\Omega}(i) = \sum_j G_{ij}^{eq}$;

\item Fixed-power Ohmic betweenness, $C_P(i)  = \sum_a \sum_{b \neq a} \frac{1}{\sqrt{G_{ab}^{eq}}}\sum_{j \neq i} I_{ij}^{ab}$, as described in \citet{thomas2009ocindac}\footnote{In brief: for all pairs of nodes $(a,b)$, a fixed power of 1 Watt is applied across the terminals corresponding to the nodes, which have an Ohmic inverse distance $G_{ab}^{eg}$. The measured current through node $i$, $\sum_{j \neq i} I_{ij}^{ab}$ determines the importance of the node to current flow between $a$ and $b$.}  (relative rank only)

\end{itemize}

Considering the absolute measures of node characteristics, it is simply a matter of calculating the statistic for each node, at each threshold, within each replicate, and converting the estimate into the units of the valued graph. The optimal threshold is that which gives the lowest squared deviation across iterations

It may also be preferable to consider only the relative importance of nodes, thereby removing the concern of a change in units. As a frequently asked question of networked systems is ``Who is the most important individual'' by some set of criteria, rank-order statistics are a logical choice to measure the change of importance of individuals between two instances of a graph. Since there is also far more interest in the more important individuals (those with rank $R_i$ closer to 1) than the less important ones (with rank $R_i$ closer to $N$), a rank discrepancy statistic of the form

\[ D_{ab} = \frac{1}{N}\sum_i \frac{(R_{ai}-R_{bi})^2}{\sqrt{R_{ai}R_{bi}}} \]

\noindent is used, where $R_{ai}$ and $R_{bi}$ are the ranks of individual $i$ in instances labelled $a$ and $b$. Ties in rank are randomly assorted so that, among other factors, an empty or complete graph is uninformative as to the supremacy of one node over another.

\subsection{Results on Simulated Data}

\begin{figure}
\begin{center}
$\begin{array}{cc}\includegraphics[width=0.5\linewidth]{\imloc 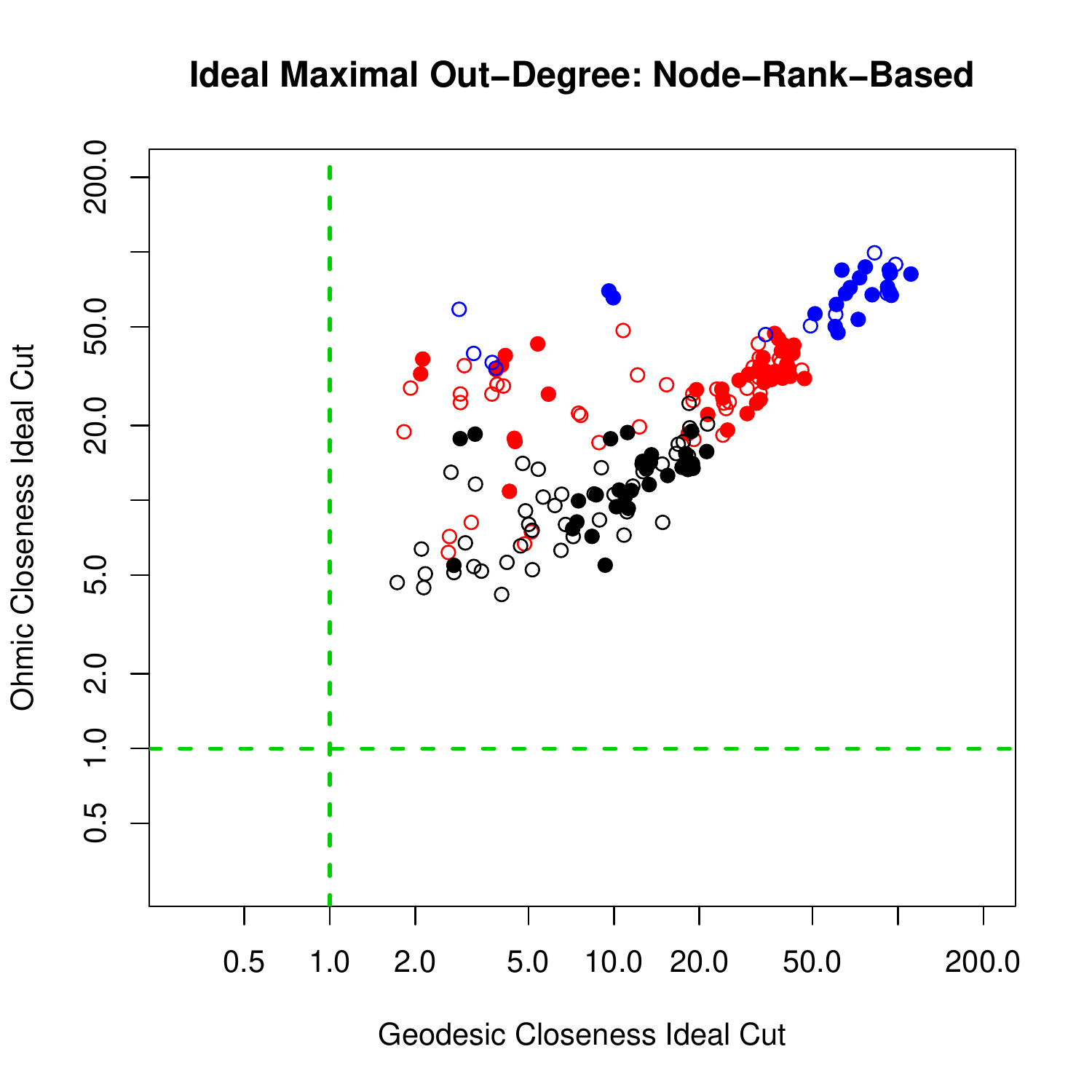} & \includegraphics[width=0.5\linewidth]{\imloc 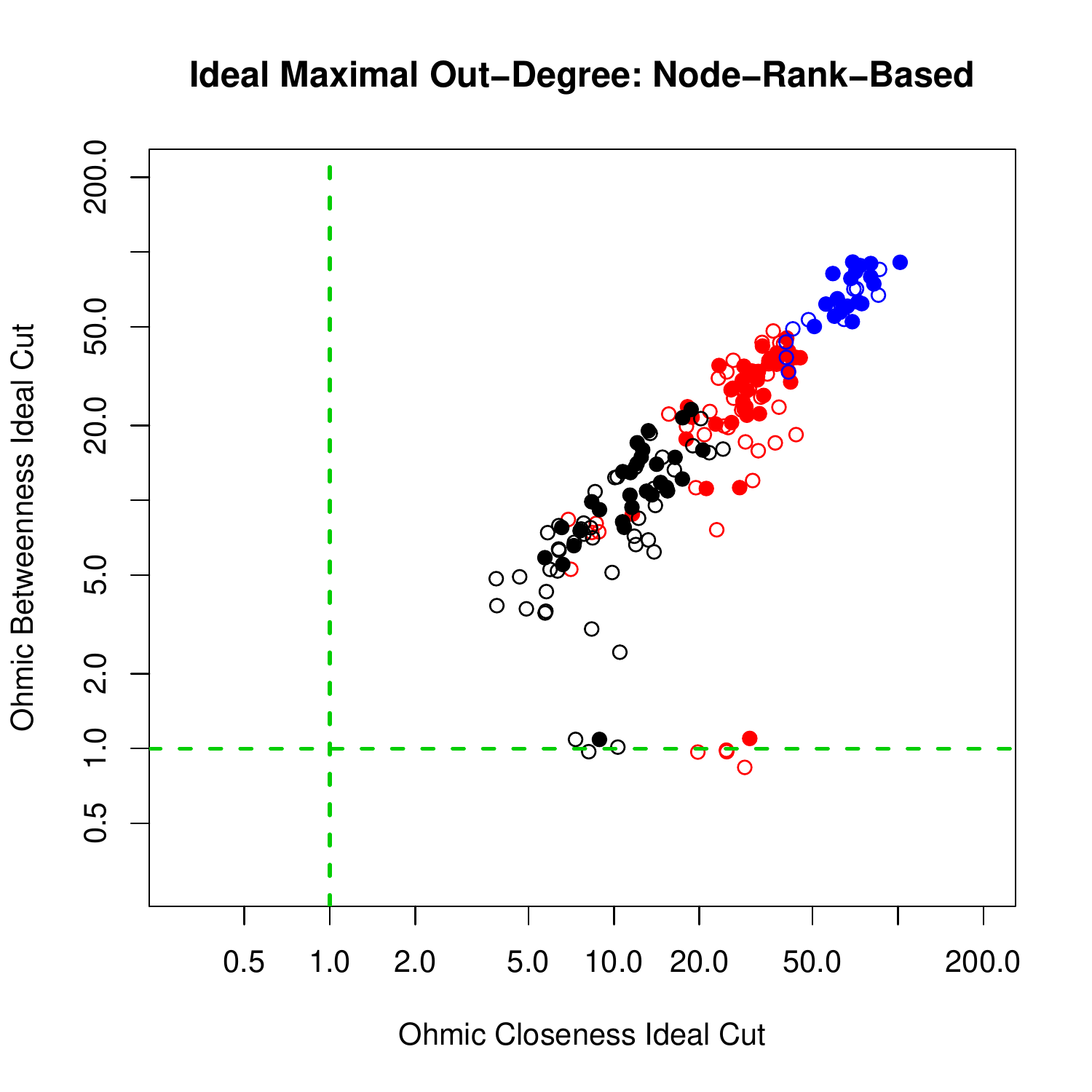} 
\end{array}$
\includegraphics[width=0.5\linewidth]{\imloc 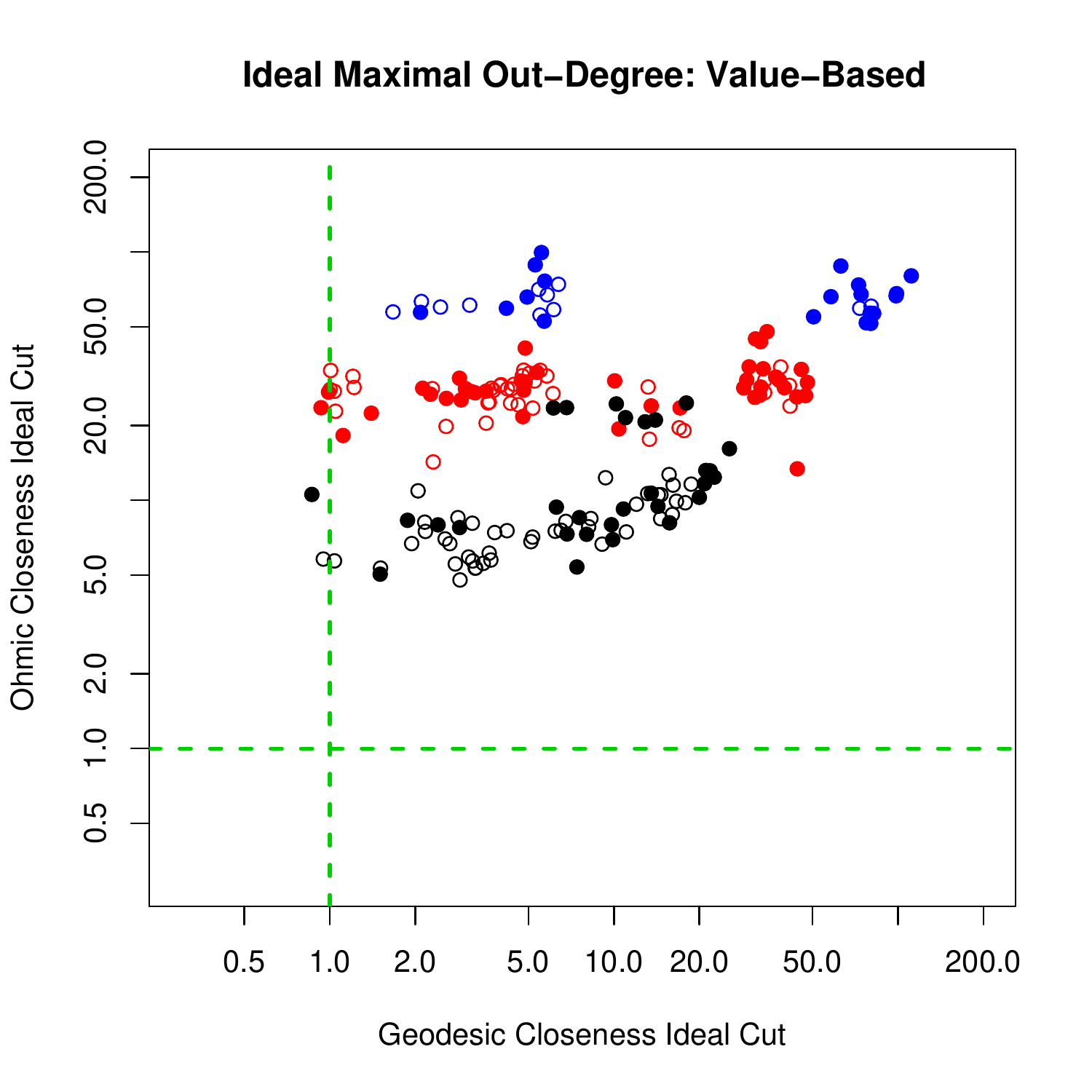}
\end{center}
\caption{\label{censor-scatter} Top: Scatterplots of ideal threshold points based on rank discrepancy statistics for harmonic geodesic centrality, Ohmic closeness and Ohmic betweenness. Note the separation of ideal cuts based on network size; black, red, blue represent networks of size 50, 100 and 200 respectively. Bottom, scatterplots of ideal threshold points based on direct value comparison for geodesic and Ohmic closeness. There is separation of the optimal cut point based on network size for Ohmic closeness, though not for geodesic closeness, suggesting that higher tie densities are required to fully capture the essence of multiple paths that is captured in Ohmic statistics.}
\end{figure}

Figure \ref{censor-scatter} shows the ideal cutpoints for a series of 183 graphical families with varying characteristics. For rank statistics, once again there is a general agreement on the ideal cutpoint by rank discrepancies, as most values lie on the diagonal of equality; note once again that in several cases in the first diagram, higher outdegree maxima are required for Ohmic statistics to be satisfied.

In all cases, there is a strong effect of network size on the ideal outdegree; restated, however, this suggests that the ideal density (edges over total edges, $n(n-1)/2$) is roughly identical for each network size in the Ohmic family of statistics. This is not the case for geodesic statistics, however, as there is a strong variability on the ideal outdegree, namely for value-based statistics. The maximum, however, does appear to scale with network size in this case.

\begin{figure}
\begin{center}
\includegraphics[width=0.8\linewidth]{\imloc 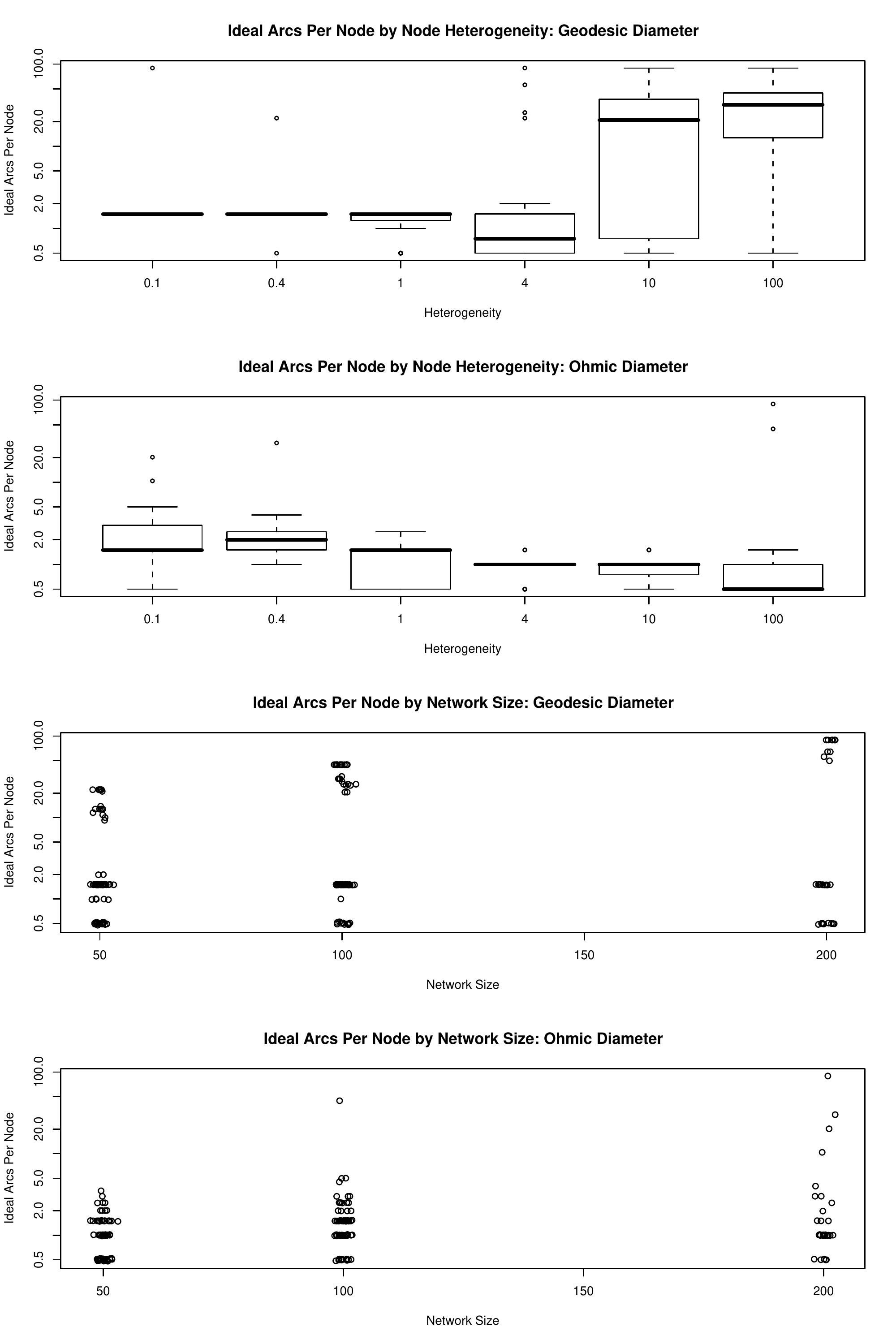}
\end{center}
\caption{\label{censor-geometry} A comparison of the ideal arcs-per-node for the geodesic and Ohmic diameters, as plotted by node heterogeneity and network size respectively. For cases of extreme nodal heterogeneity, geodesic diameter is optimized at much higher out-degree than in the Ohmic case. Similarly, when plotting outcomes as a function of network size, there are far more points of geodesic optimality that lie at the extremes (average outdegree, (0.5, 1.0, 1.5) as well as around N/2) as opposed to Ohmic distance measures.} 
\end{figure}

Figure \ref{censor-geometry} details the optimal maximal outdegrees for diameter when dividing the cases by generative parameter. In this case, two inputs proved to show the most discrimination between optima: heterogeneity on degree and network size. First, there is a preference for denser graphs for the geodesic case, and not the Ohmic, the opposite scenario when considering closeness statistics. Additionally, it is the more heterogeneous cases that drive this behaviour; much greater inclusion is necessary as heterogeneity increases beyond $\sigma_{\alpha}=4$. This behaviour is independent of network size, as the extreme values are present for geodesic diameter at each choice of size.

\subsection{Results on Real Data Examples}

This method of dichotomization was chosen largely as a convenient alternative to the straight thresholding method, but its value is best proven on examples of real data that may need the method to be analyzed. In particular, we consider two data sets:

\begin{itemize}

\item The EIES electronic communications data of \citet{freeman1980scsesng}, in which arcs from one of 32 individuals to another is a message count. There are many arcs with value equal to zero. 

\item The fMRI brain-wave data of \citet{achard2006rls} in which the 90 nodes correspond to brain regions and the edges are, essentially, partial correlations of signals. In contrast, there are no zeroes in this data if we consider (for demonstration's sake) the partial correlations to be without error.

\end{itemize}

The results for the straight thresholding procedure were presented graphically in \citet{thomas2010vttfl}.

\begin{table}
\begin{center}
\begin{tabular}{c|cccc}
\hline
\hline
\textbf{Freeman-Freeman EIES} & Thresholding & Thresholding & Censoring & Censoring \\
Measure & Arcs/Node & Optimum & Optimum & Outdegree \\
\hline
Geodesic Centrality Rank & 6 & \textbf{24.5} & 51.97 & 18 \\
Ohmic Centrality Rank & 7 & \textbf{7.79} & 35.69 & 24+ \\
Ohmic Betweenness Rank & 17 & \textbf{6.445} & 7.807 & 24+ \\
Geodesic Centrality Value & 6 & \textbf{6715} & 18160 & 24+ \\
Ohmic Centrality Value & 11 & \textbf{16480} & 59530 & 11 \\
Geodesic Diameter & 15 & \textbf{0.09056} & 0.942 & 10 \\
Ohmic Diameter & 17 & 0.4646 & \textbf{0.4131} & 12 \\
\hline
\hline
\end{tabular}
\end{center}
\caption{Comparing the optimal dichotomizations for the 32-node EIES network \citep{freeman1980scsesng} under straight thresholding and the deliberate censoring of outdegree. In five of the seven network measures, the thresholding method is far superior at conserving the property from the valued graph; the diameter measures yield essentially identical results. \label{table:freeman-comp}}
\vskip 2cm
\begin{center}
\begin{tabular}{c|cccc}
\hline
\hline
\textbf{Achard Brainwave} & Thresholding & Thresholding & Censoring & Censoring \\
Measure & Arcs/Node & Optimum & Optimum & Outdegree \\
\hline
Geodesic Centrality Rank & 14.72 & \textbf{149.5} & 372.1 & 18.4 \\
Ohmic Centrality Rank & 22.2 & \textbf{132.8} & 278.5 & 20 \\
Ohmic Betweenness Rank & 22.2 & \textbf{251.9} & 386.8 & 27.7 \\
Geodesic Centrality Value & 80+ & 0.3587 & 0.3609 & 81 \\
Ohmic Centrality Value & 80+ & 41.37 & 39.67 & 81 \\
Geodesic Diameter & 33.4 & 0.018 & \textbf{0.013} & 18.4 \\
Ohmic Diameter & 68.8 & \textbf{0.069} & 0.126 & 54.3 \\
\hline
\hline
\end{tabular}
\end{center}
\caption{Comparing the optimal dichotomizations for the 90-node Achard brain network \citep{achard2006rls} under straight thresholding and the deliberate censoring of outdegree. In the nodal rank measures, the straight thresholding dominates the censoring by outdegree; each method is superior at one type of diameter but not the other. The measures on centrality values, however, are largely ignorable, as both insist on networks that are essentially complete. Note that the statistics are not comparable between the EIES and Achard examples, as the scales for each measure vary both with the underlying network size and the relative tie scales.\label{table:achard-comp}}
\end{table}


As table \ref{table:freeman-comp} shows for the results for the EIES data, straight thresholding is preferred hands down; fix of the seven test statistics are significantly smaller under thresholding than censoring at their ideal levels, and the seventh (Ohmic diameter) is on roughly the same scale. This is likely because the EIES data are extremely heterogeneous in popularity/gregariousness -- some individuals sent far more communications than others, to a wider variety of people -- which is more sensitive to raw censoring than raw thresholding in the case of the most central individuals.

For the Achard data, the solution is less clear. The rank-based criteria favour the thresholded version; the diameters each prefer one method or the other; and the value-based statistics both essentially suggest a nearly complete graph, the act of censoring on outdegree may appear similar to the act of thresholding, at least from the point of view of the outdegree on each node. Still, the notion that an already questionable procedure in thresholding produces a better result than outdegree censoring in several cases is noteworthy.

\section{Effect on Linear Models}

We turn now to the use of the dichotomized graph as the measurer of contagion in a two-time-step linear model framework. For the sake of demonstration we use the same linear model formulation as was used in the thresholding analysis of \citet{thomas2010vttfl},

\[ Y_{i1} = \mu + \gamma Y_{i0} + \beta \sum_j X_{ij0} Y_{j0} + \varepsilon_{i1}, \]

\noindent so that the model is generated by the original valued graph and estimated using either the valued or degree-censored cases. The performance in estimating the contagion effect $\beta$ is measured by taking the mean squared error of the estimates against the generated value.   

\begin{figure}
\begin{center}
\includegraphics[width=0.7\linewidth]{\imloc 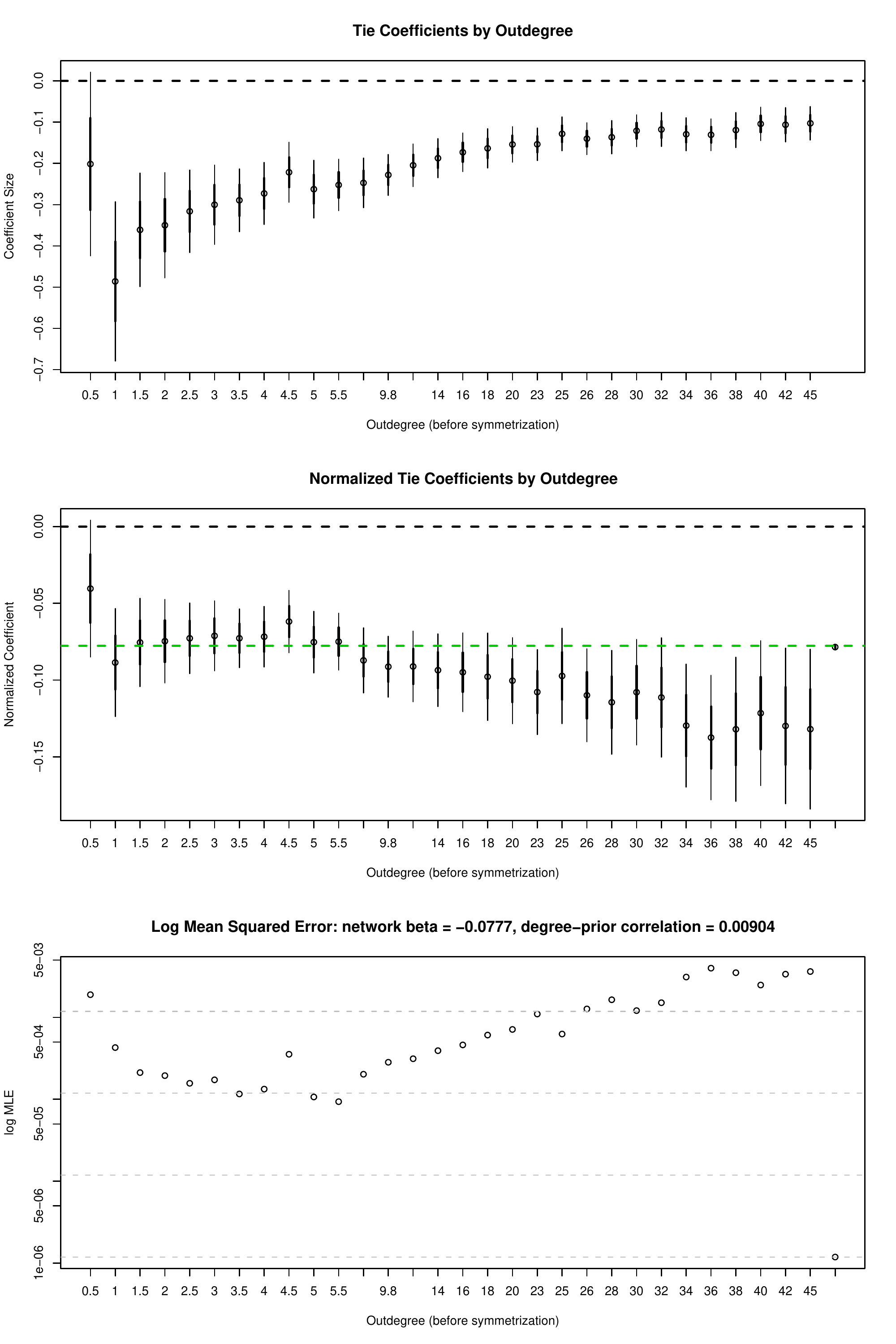}
\end{center}
\caption{A censoring profile for a single valued graph, where directed ties have been symmetrized. In the best case, the mean squared error is 100 times larger for a dichotomized network than the valued case.\label{cens-lm-sym-1}}
\end{figure}

Figure \ref{cens-lm-sym-1} shows a single instance of censoring by outdegree for a linear model at various maximal outdegrees for the network parameter $\beta$. The example in this case has a negligible correlation between nodal indegree and past property value, and while the value at one out-arc per node is inconsistent with the true generative parameter, many of the rest come quite close in their estimates adjusted for scale. Still, the inflation in the estimate's mean squared error is noticeably large, 100 times larger than the (true) valued case.

\begin{figure}
\begin{center}
\includegraphics[width=\linewidth]{\imloc 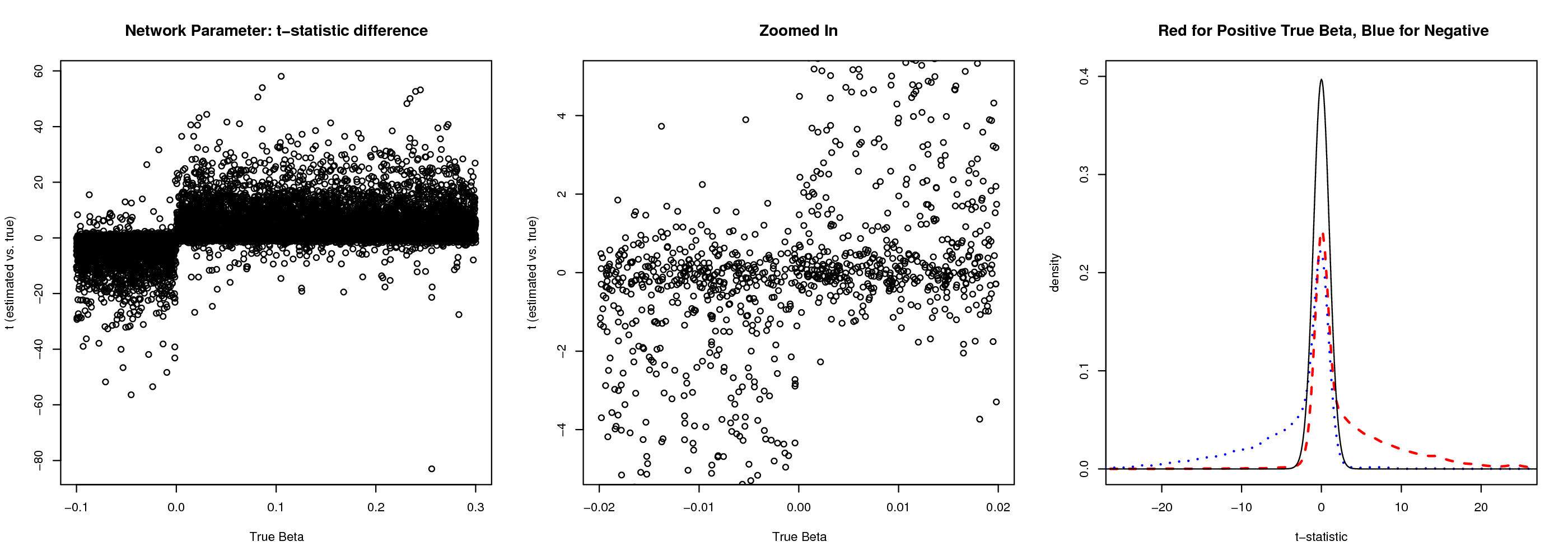}
\end{center}
\caption{The $t$-statistics for the estimates of the autocorrelation term $\beta$, with respect to the true generated value. By conditioning on the true value, it becomes clear that the estimates are inflated in magnitude with respect to the truth.\label{beta-split-cens}}
\end{figure}

In the aggregate, however, the same issues persist in this compression mode: estimates for $\beta$ are highly inflated in absolute value under most conditions, as seen in Figure \ref{beta-split-cens} and as was observed in the thresholding case.

\begin{figure}
\begin{center}
\includegraphics[width=0.7\linewidth]{\imloc 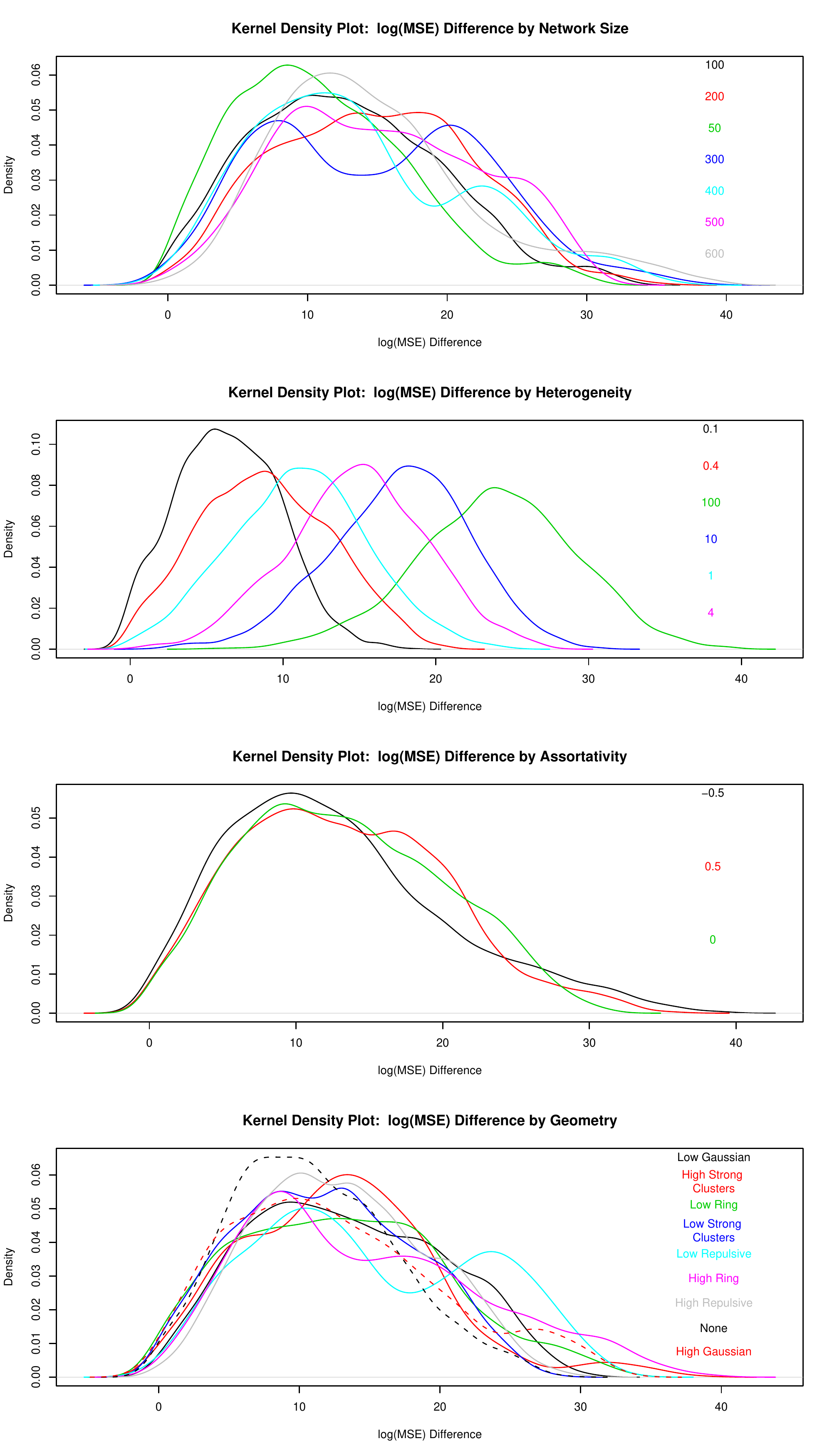}
\end{center}
\caption{\label{cens-collective-sym} Scatterplot and density plot of the minimal mean squared error ratio for thresholding tests. The x-axis of the scatterplot is the generated correlation between indegree and the prior outcome value, and no effect is visible.}
\end{figure}

When considering aggregate loss of efficiency in terms of mean squared error ratio for $\beta$, the same suspect appears to be the most discriminating. Heterogeneity by node popularity is easily the most influential in determining the inefficiency in these estimates, as shown in Figure \ref{cens-collective-sym}. Even small differentiation provides a major shift in precision, with a rough increase in scale of $e^5 \approx 150$ times the mean squared error in the valued case. None of the other generative parameters have this level of influence on the efficiency of the censoring-based estimation.

\section{Conclusions}

The preceding investigation of deliberate censoring is worthwhile, in that it shows censoring by outdegree appears to be just as lossy for network compression as thresholding, even when changes of scale are factored in. However, this in itself is not the end of the story; while both censoring and thresholding are methods that can be used to construct binary approximations to valued graphs, minimizing a chosen statistic is an approach that transcends both censoring and thresholding. In particular, the ideal solution to the dichotomization issue would be to search the space of all $2^{n(n-1)}$ binary graphs and select the graph that best approximates the valued graph. In this way, both thresholding and censoring produce excellent starting points for randomized searches and hill-climbing approaches.

In both investigations, the biggest factor for information loss is the heterogeneity of node popularity. Since this factor affects both the number and strength of nonzero edges, it is entirely likely that the distortion is due to one or both of these effects. A generative model that could investigate this trend would do well to separate the two in some way. In particular, consider the situation where one person has to divide the hours in a day between the people in the rest of the social environment. The outcome here would stretch between two alternatives: many shallow interactions (casual friends) and one strong interaction (a best friend or partner).

Incorporating this into a model places emphasis on variability over mean value. A plausible mechanism to generate this scenario would incorporate this heterogeneity into variability, such as in the normal form

\begin{eqnarray}
Z_{ij} & = & \mu + \varepsilon_{ij}; \\
\varepsilon_{ij} & \sim & N(0, \alpha_i \alpha_j); \\
\alpha_i & \sim & \frac{1}{c} Gamma(c),
\end{eqnarray}

\noindent which has have two driving parameters $\mu$ and $\varepsilon$, though the first would be, in this proposal, uniform across all edges.


That said, the main objection remains: there is still a considerable loss of information, and an introduction of bias, that takes place when this operation is conducted. With any loss of statistical power, it still cannot be guaranteed that a compressed graph structure will produce estimates with the correct coverage properties, considerably jeopardizing the scientific value of the particular investigation.

\bibliographystyle{\baseloc ims}
\bibliography{\baseloc actbib}

\begin{thebibliography}{11}
\expandafter\ifx\csname natexlab\endcsname\relax\def\natexlab#1{#1}\fi
\expandafter\ifx\csname url\endcsname\relax
  \def\url#1{\texttt{#1}}\fi
\expandafter\ifx\csname urlprefix\endcsname\relax\def\urlprefix{URL }\fi
\providecommand{\eprint}[2][]{\url{#2}}

\bibitem[{Achard et~al.(2006)Achard, Salvador, Whitcher, Suckling and
  Bullmore}]{achard2006rls}
\textsc{Achard, S.}, \textsc{Salvador, R.}, \textsc{Whitcher, B.},
  \textsc{Suckling, J.} and \textsc{Bullmore, E.} (2006).
\newblock {A Resilient, Low-Frequency, Small-World Human Brain Functional
  Network with Highly Connected Association Cortical Hubs}.
\newblock \textit{The Journal of Neuroscience}, \textbf{26} 63--72.
\newblock \\\urlprefix\url{http://www.jneurosci.org/cgi/content/full/26/1/63}.

\bibitem[{Brandes and Fleischer(2005)}]{brandes2005cmbcf}
\textsc{Brandes, U.} and \textsc{Fleischer, D.} (2005).
\newblock {Centrality Measures Based on Current Flow}.
\newblock In \textit{22nd Symposium on Theoretical Aspects of Computer Science
  (STACS’05)}. 533–544.

\bibitem[{Freeman and Freeman(1980)}]{freeman1980scsesng}
\textsc{Freeman, L.} and \textsc{Freeman, S.} (1980).
\newblock {A Semi-Visible College: Structural Effects on a Social Networks
  Group}.
\newblock In \textit{Electronic Communication: Technology and Impacts}.
  Westview Press, 77--85.

\bibitem[{Freeman(1979)}]{freeman1979csncc}
\textsc{Freeman, L.~C.} (1979).
\newblock {Centrality In Social Networks: Conceptual Clarification}.
\newblock \textit{Social Networks}, \textbf{1} 215--239.

\bibitem[{Granovetter(1973)}]{granovetter1973swt}
\textsc{Granovetter, M.} (1973).
\newblock {The Strength of Weak Ties}.
\newblock \textit{American Journal of Sociology}, \textbf{78} 1360--1380.

\bibitem[{Klein and Randic(1993)}]{klein1993rd}
\textsc{Klein, D.} and \textsc{Randic, M.} (1993).
\newblock {Resistance distance}.
\newblock \textit{Journal of Mathematical Chemistry}, \textbf{12} 81--95.

\bibitem[{Newman(2005)}]{newman2005mbcbrw}
\textsc{Newman, M.} (2005).
\newblock {A Measure of Betweenness Centrality Based on Random Walks}.
\newblock \textit{Social Networks}, \textbf{27} 39–54.

\bibitem[{Thomas(2009)}]{thomas2009ocindac}
\textsc{Thomas, A.~C.} (2009).
\newblock {Ohmic Circuit Interpretations of Network Distance and Centrality}.
\newblock Unpublished manuscript.,
  \\\urlprefix\url{http://www.acthomas.ca/academic/relational.htm}.

\bibitem[{Thomas(2010)}]{thomas2010cocisnpeaa}
\textsc{Thomas, A.~C.} (2010).
\newblock {Censoring Outdegree Compromises Inferences of Social Network Peer
  Effects and Autocorrelation}.
\newblock Submitted to Sociological Methodology,
  \\\urlprefix\url{http://arxiv.org/abs/1008.1636}.

\bibitem[{Thomas and Blitzstein(2010{\natexlab{a}})}]{thomas2010mshmfrd}
\textsc{Thomas, A.~C.} and \textsc{Blitzstein, J.~K.} (2010{\natexlab{a}}).
\newblock {Marginally Specified Hierarchical Models for Relational Data}.
\newblock Unpublished manuscript,
  \\\urlprefix\url{http://www.acthomas.ca/papers/a-framework-for-modelling.pdf%
}.

\bibitem[{Thomas and Blitzstein(2010{\natexlab{b}})}]{thomas2010vttfl}
\textsc{Thomas, A.~C.} and \textsc{Blitzstein, J.~K.} (2010{\natexlab{b}}).
\newblock {Valued Ties Tell Fewer Lies: Why Not To Dichotomize Network Edges
  With Thresholds}.
\newblock Submitted to Annals of Applied Statistics.

\end{thebibliography}

\end{document}